\documentclass[journal]{IEEEtran}

\normalsize

\usepackage{multirow}
\usepackage[dvips]{graphicx}
\usepackage{amsbsy}	
\usepackage[keeplastbox]{flushend}

\begin{document}

\title{A Synthetic Statistical MIMO PLC Channel Model Applied to an In-Home Scenario}

\author{Alberto~Pittolo~and~Andrea~M.~Tonello~
\thanks{Alberto Pittolo is with the Polytechnic Department of Engineering and Architecture (DPIA), University of Udine, Udine 33100, Italy e-mail: \mbox{alberto.pittolo@uniud.it}.}
\thanks{Andrea M. Tonello is with the Institute of Networked and Embedded Systems, Alpen-Adria-Universit\"{a}t Klagenfurt, Klagenfurt 9020, Austria, email: \mbox{andrea.tonello@aau.at}.}}

\markboth{\hfill A version of this paper has been accepted for publication to IEEE Transactions on Communications -- \textcopyright\ IEEE 2017 \hfill}{}

\newcommand{\figurewidth}{\columnwidth}
\newcommand{\tabstretch}{1.2}

\maketitle

\begin{abstract}
This paper proposes a synthetic statistical top-down MIMO power line communications channel model based on a pure phenomenological approach. The basic idea consists of directly synthesizing the experimental channel statistical properties to obtain an extremely compact model that requires a small set of parameters. The model is derived from the analysis of the in-home $2\times3$ MIMO PLC channel data set obtained by the ETSI Specialist Task Force 410 measurement campaign in the band 1.8--100 MHz. The challenge of modeling the channel statistical correlation, exhibited among the frequencies and between the MIMO modes, in compact form is tackled and it is shown that a small set of parameters can be used to reconstruct such a correlation behavior. The model is validated and compared to the measured channels, showing a good agreement in terms of average channel gain, root-mean-square delay spread, coherence bandwidth, and channel capacity distribution.
\end{abstract}

\begin{IEEEkeywords}
Power line communications, PLC, channel modeling, MIMO, phenomenological approach, top-down methodology, statistical model.
\end{IEEEkeywords}

\IEEEpeerreviewmaketitle

\section{Introduction}
\IEEEPARstart{R}{ecently}, the exploitation of multiple-input multiple-output (MIMO) transmission techniques has become of great interest to support the massive demand for high-speed data services in communication networks. Although MIMO techniques have been extensively investigated within the wireless (radio) scenario, they have been only recently analyzed in the context of power line communications (PLC). In practice, when an electric power distribution network deploys multiple conductors, PLC can exploit MIMO transmission by injecting and receiving signals among the multiple wires connecting the nodes \cite{MIMObook}. This is the case, for instance, of in-home low voltage (LV) power distribution networks that deploy three conductors: the phase, the neutral and the protective earth wires. The recent Home Plug AV2 system \cite{HPAV2} and the ITU G.hn standard \cite{ITU_Ghn} have adopted MIMO transmission together with a bandwidth extension (from 2 to 86 MHz), precoding, parameters adaptation and a number of advanced techniques at the MAC layer that enable high data rates in excess of 1 Gbps of peak rate. This impressive performance level stimulates the interest in PLC solutions, their evolution and their application from in-home networking to  smart grid applications \cite{LampeTonello_book,SmartGrid2030}.

In order to develop algorithms and assess performance, it is very important to characterize and model the MIMO PLC channel. Single-input single-output (SISO) PLC channel modeling is a rather more established area of research than MIMO PLC channel modeling \cite[Ch.~2]{LampeTonello_book}. Essentially, two different approaches can be followed: a deterministic approach, which assumes a specific network topology, and a statistical approach, where some variability is introduced to produce random channel realizations. The statistical method is more suitable for a comprehensive model since it statistically emulates the variability encountered within a certain scenario. Contrariwise, a deterministic method well represents a specific configuration, but with a poor overall representativeness. Both these approaches can then be specialized into a bottom-up or a top-down methodology.

The bottom-up methodology has a tight connection with the physical propagation effects, enabling a faithful emulation of an individual channel frequency response (CFR) by exploiting the transmission line theory. Its application requires a full network topology knowledge in terms of cable characteristics, layout, connection lengths and loads, resulting into a considerable computational effort. An example of a bottom-up deterministic SISO PLC channel model can be found in \cite{Galli_06}. A statistical bottom-up channel model can then be obtained by using a random topology model as shown in \cite{BotUpMod_p1}.

Contrariwise, the top-down methodology uses an analytic function, whose parameters are chosen to fit an actual channel measurement in an empirical way. The top-down approach facilitates the statistical extension with respect to (w.r.t.) the bottom-up approach. One of the first and well-known top-down approaches was discussed in \cite{PLC_MPM}, where the CFR was modeled taking into account the multipath signal propagation and the cable losses. Later, it was shown that, by introducing variability in some parameters of the fitting function, the model in \cite{PLC_MPM} can be statistically extended \cite{Tonello_12}.

Although much effort has been spent in providing faithful SISO models, summarized in \cite[Ch.~2]{LampeTonello_book}, only few proposals have been made to define a general MIMO PLC channel model so far. This is also due to the fact that MIMO PLC channel measurement and characterization has started only recently. The most comprehensive in-home measurement campaign has been conducted by the European Telecommunication Standards Institute (ETSI) Specialist Task Force 410 (STF-410) \cite{ETSI1, ETSI3} in 2012. A first work that extends the SISO bottom-up model in \cite{BotUpMod_p1} to the MIMO case by using multi-conductor transmission line theory based on a large measurement database was described in \cite{Versolatto_11}. Very recent improvements have been proposed in \cite{CortesJSAC}. Conversely, a first attempt to model the MIMO PLC channel impulse response via a top-down approach was detailed in \cite{Canova_10}, while another frequency domain model was proposed in \cite[Ch.~5]{MIMObook} and recently improved in \cite{PaganiJSAC}. These MIMO models start from an analytic fitting function similar to that proposed in \cite{PLC_MPM} and then they render some parameters random similarly to the SISO approach in \cite{Tonello_12}.

The main contribution of this paper is to show that a synthetic top-down approach, based on a pure phenomenological philosophy, can be followed to provide a simple model that is statistically representative of a real wideband MIMO PLC environment. Due to the model baseline methodology, it can be potentially applied to other communication scenarios, since it is not connected to any physical assumption or propagation phenomena. Furthermore, its underlying concept significantly differs from what has been already proposed in the literature. Indeed, differently from the typically more complex transmission line bottom-up methods, e.g. \cite{CortesJSAC}, or the conventional top-down approaches that use a fitting function related to some physical assumptions, e.g. \cite{PaganiJSAC}, the proposed statistical model is purely phenomenological and directly synthesizes the MIMO CFR and its statistical properties. For this reason, an initial characterization part is fundamental to first assess the MIMO channel properties and then show how to reproduce them by using a reduced set of parameters. This is done by considering all the measurements of a specific database, obtained from the in-home measurement campaign in the band 1.8--100 MHz, carried out by the ETSI STF-410 \cite{ETSI1,ETSI3}.

The model is named ``synthetic'' since this terminology encompasses two meanings, namely ``artificial'', i.e. obtained from synthesis without physical assumptions (phenomenological), and ``compact'', since the model adopts a small number of parameters. The theoretical formulation resembles the top-down strategy that we presented in \cite[Sec.~3.1]{IEICE_14} for the in-home SISO channel, but it is extended to the far more complex MIMO scenario. Moreover, the proposed model adopts a completely new strategy to take into account both the frequency and the MIMO ports (or modes) correlation, represented by the MIMO statistical covariance matrix. In particular, the proposed synthetic model implementation strategy is able to reconstruct the overall MIMO matrix by using a significantly reduced number of parameters. Conversely, the method based on the MIMO extension of \cite{IEICE_14}, despite its theoretical simplicity, must have available all the matrix coefficients and adopts a different and more complex strategy for the phase generation. This is why it is referred to as not-fully synthetic in the rest of the paper.

The model philosophy is first introduced in Section~\ref{sec:ModelBasics}. The experimental data is analyzed in Section~\ref{sec:ExprEvid}, enabling the derivation of the channel properties, such as the statistical behavior and the exhibited statistical correlation among the frequencies and between the MIMO modes. Then, the synthetic modeling strategy, which uses a reduced set of parameters to reconstruct the MIMO PLC channel characteristics, is detailed and justified by the experimental evidence in Section~\ref{sec:ModelDescript}. This description guarantees the methodology replicability. The results are validated in Section~\ref{sec:Results} by showing that the proposed statistical model is capable to replicate the measured channels in terms of statistical metrics, i.e. average channel gain (ACG), root-mean-square delay spread (RMS-DS), and coherence bandwidth (CB) \cite{TCOMp1}, as well as in terms of MIMO modes correlation measured by the condition number $\kappa$ \cite{CortesJSAC}. The channel capacity distribution (or maximum achievable rate according to Shannon's definition) is also considered as a further benchmarking tool. All the mentioned quantities have been computed considering the overall measurements database. A comparison with the not-fully synthetic procedure, as well as with the results provided by the synthetic strategy, when considering a reduced set of MIMO modes, is also performed within this section. Finally, the conclusions follow in Section~\ref{sec:conc}.

\section{Model Philosophy}
\label{sec:ModelBasics}
Most of the proposed top-down channel models, e.g. \cite{PLC_MPM}, are based on an analytic function that represents some physical phenomena, such as the multipath propagation due to the several branches and load mismatches that cause multiple signal reflections in a PLC network. Similarly, the model presented in \cite{PaganiJSAC} is based on the theoretical formulation provided in \cite{PLC_MPM}. Therefore, some physical assumptions are made, resulting into a hybrid top-down and bottom-up model. Contrariwise, the basic idea underlying the proposed model is fully phenomenological and focuses on a pure top-down procedure.

To start, the generic SISO CFR at frequency $f$, namely $H(f)$, is a complex number that can be expressed in amplitude $A(f)$ and phase $\varphi(f)$, i.e. $H(f)=A(f)e^{i\varphi(f)}$. In order to simplify the discussion, the CFR in dB scale, namely $H_{dB}(f)=20\log_{10}{H(f)}$, is considered, yielding
\begin{equation}
\label{eq:CFRdB}
  H_{dB}(f)=A_{dB}(f)+i K\varphi(f),
\end{equation}
where the real part $A_{dB}(f)=20\log_{10}|H(f)|$ is the amplitude in dB scale, while the imaginary part is still the phase in radians multiplied by the constant value $K=20\log_{10}e$.

In practice, when considering a PLC scenario, the quantity $H(f)$ changes depending on the considered network topology and the specific pair of nodes, e.g., outlets in a home network. Thus, from a statistical perspective, $H(f)$ can be modeled as a random variable (RV), while the ensemble of all the available $N_f$ frequency samples can be grouped into a channel vector of $N_f$ RVs. Similarly, a general MIMO channel can be represented by a three-dimensional (3D) matrix $\mathbf{H}$ of RVs with dimensions $N_R \times N_T \times N_f$, where $N_R$, $N_T$ are the number of receiving and transmitting ports, respectively.

It follows that a MIMO PLC channel is completely described by an equivalent 3D matrix of RVs characterized by certain statistics and relationships. Therefore, a typical communication scenario can be emulated by generating a set of simulated channels having a statistical behavior and, in particular, a frequency and MIMO modes correlation that are equivalent to those derived from experimental data. Hence, to provide an effective channel emulator, real data must be analyzed to propose a replicable modeling approach, possibly with a reduced set of parameters. This is the subject of the next sections.

\label{ssec:ThForm}
The proposed model considers the marginal distribution of the CFR, i.e. the distribution function of the amplitude and phase, as well as the their statistical correlation between the different MIMO modes and among the different frequencies. However, as it will be clarified in Section~\ref{sec:ExprEvid}, the CFR amplitude and phase can be approximated as independent. Thus, their marginal distribution and statistical auto-correlation are sufficient to provide an order two statistics of the channel.

In order to obtain such statistics from real data and to simplify the modeling procedure, the 3D matrix $\mathbf{H}$ is reshaped into a vector $\widetilde{\mathbf{H}}$ of size $M=N_R \cdot N_T \cdot N_f$. Thus, the first $N_f$ elements concern all the frequency samples of the first receiving and transmitting port. The second $N_f$ elements correspond to the frequencies of the second receiving port and the first transmitting port, and so on for all the $N_R$ receiving ports. Afterwards, all the other $N_T$ transmitting modes are considered accordingly. Hence, the 3D matrix $\mathbf{H}$ is reshaped as
\begin{equation}
\label{eq:ReshMtx}
 \widetilde{\mathbf{H}}=[H_{1,1}(\mathbf{f}),\dots,H_{1,N_R}(\mathbf{f}),H_{2,1}(\mathbf{f}),\dots,H_{N_T,N_R}(\mathbf{f})]^T,
\end{equation}
where $\{\cdot\}^T$ denotes the transpose operator, while $H_{i,j}(\mathbf{f})$ is the row vector of RVs associated to the frequency samples vector $\mathbf{f}$ for the CFR between the $i$-th transmitting and the $j$-th receiving ports.

The statistical covariance matrix, of size $M \times M$, of the reshaped vector $\widetilde{\mathbf{H}}$ is given by
\begin{equation}
\label{eq:CovMtx}
  \mathbf{Q}(p,q)=E\left[\left(\widetilde{H}_{i,j}(n)-\eta_{i,j}(n)\right)\left(\widetilde{H}_{\ell,r}(m)-\eta_{\ell,r}(m)\right)^*\right],
\end{equation}
where $\{\cdot\}^*$ is the complex conjugate operator and $E[\cdot]$ is the expectation operator w.r.t. the overall set of realizations. Instead, $\widetilde{H}_{i,j}(n)$ is the CFR between the $i$-th transmitter and the $j$-th receiver port at the $n$-th frequency sample, with $i=1,\dots,N_T$, $j=1,\dots,N_R$ and $n=1,\dots,N_f$. The quantity $\eta_{i,j}(n)=E[\widetilde{H}_{i,j}(n)]$ is the corresponding average value. Moreover, $p$ and $q$ are the row and column indexes of $\mathbf{Q}$ that, according to the previously described reshaping method, are function of $(n,i,j)$ and $(m,\ell,r)$, respectively. In particular, the row index is given by $p=(i-1)N_RN_f+(j-1)N_f+n$, and the same applies to the column index $q$ for the values $m$, $\ell$ and $r$.

In the following, the normalized covariance matrix $\mathbf{R}(p,q)=\mathbf{Q}(p,q)/(\sigma_p\sigma_q)$ is also considered. It is computed as the covariance matrix in (\ref{eq:CovMtx}) normalized by the product of the related RVs standard deviation, i.e. $\sigma_p$ and $\sigma_q$. Each element of $\mathbf{R}$ ranges from 0 (no correlation) to 1 (fully correlated) and thus it is equivalent to the Pearson correlation coefficient \cite{Gili_11}. Hence, each coefficient of the matrix $\mathbf{R}$ represents the correlation degree among the corresponding frequency samples and between the different transmitting and receiving ports (MIMO modes correlation).

By exploiting (\ref{eq:CFRdB}) and (\ref{eq:CovMtx}), the covariance matrix, of size $M \times M$, of the CFR in dB can be obtained as
\begin{equation}
\label{eq:CovDep}
  \mathbf{Q}_{H_{dB}}=\mathbf{Q}_{A_{dB}}+K^2\mathbf{Q}_\varphi+i K\left[\mathbf{Q}_{A_{dB},\varphi}^T-\mathbf{Q}_{A_{dB},\varphi}\right],
\end{equation}
where $\mathbf{Q}_{A_{dB},\varphi}$ is the covariance matrix between the amplitude (in dB scale) and the phase, while $\mathbf{Q}_{A_{dB}}$ and $\mathbf{Q}_\varphi$ are the auto-covariance matrices of the amplitude in dB and the phase, respectively. Obviously, in this case, the amplitude and the phase are vectors reshaped as in (\ref{eq:ReshMtx}), equivalently to the CFR vector $\widetilde{\mathbf{H}}$.

As it can be noted, the real part of $\mathbf{Q}_{H_{dB}}$ depends only on the amplitude and phase auto-covariance matrices, while the imaginary part is a function of the mutual covariance matrix, as discussed in \cite{ProperComplex}. The knowledge of the covariance matrices $\mathbf{Q}_{H_{dB}}$, $\mathbf{Q}_{A_{dB}}$, $\mathbf{Q}_\varphi$ and $\mathbf{Q}_{A_{dB},\varphi}$ provide information about the joint distribution function of the CFR amplitude and phase, which gives a second order characterization of the channel in a certain frequency range, as detailed in \cite{ProperComplex}. Hence, according to the above formulation, only the amplitude and phase properties need to be assessed and reproduced to obtain the desired behavior, providing an extremely compact procedure, named synthetic modeling strategy.

\section{Experimental Evidence}
\label{sec:ExprEvid}
In this section, a database of MIMO PLC in-home experimental CFRs is considered to obtain the statistics needed by the model, whose basic elements were described in the previous section. The MIMO in-home channel scenario has relevant application value. The measurements are briefly discussed first, assessing the main statistical properties and the relationships required for the generation process. This analysis highlights several properties that enable further simplifications to the general model, as detailed in Section~\ref{sec:ModelDescript}. 

\subsection{Measurement Campaign}
\label{ssec:MisCamp}
The database of experimental channels collected during the measurement campaign across Europe by the ETSI STF-410 \cite{ETSI1} is herein evaluated. The database consists of 353 MIMO CFRs with 1588 samples in the extended 1.8--100 MHz frequency range measured via a vector network analyzer.  The in-home LV power distribution networks herein considered deploy three wires, two for the power supply, namely the phase (P) and the neutral (N), and one for protection in case of an insulation fault, i.e. the protective earth (E). PLC modems can instantiate MIMO communication as follows \cite{MIMObook}. At the transmitter side a differential \mbox{$\Delta$-style} transmission can be implemented by signaling among a pair of wires. At the receiver side, instead, the signal can be observed between each wire and a reference plane in a star-style configuration. Beyond the three standard star-style receiving modes, an additional mode, named common mode (CM), can be extracted \cite{ETSI1}. Due to Kirchhoff's law, only two out of three transmitting voltages are linearly independent. The third is a linear combination of the other two. Thus, only two signals can be injected. The same thing applies at the receiver, i.e., among the four acquired signals, only three are linearly independent (see \cite{ETSI1} for details). Hence, the $2\times3$ MIMO channel that includes the PN and PE \mbox{$\Delta$-style} transmitting ports and the P, N, and CM star-style receiving ports, is considered in the following.

\subsection{MIMO Channel Properties}
\label{ssec:ExpChPropMODEL}
\begin{figure}[t]
\centering
\includegraphics[width=\figurewidth]{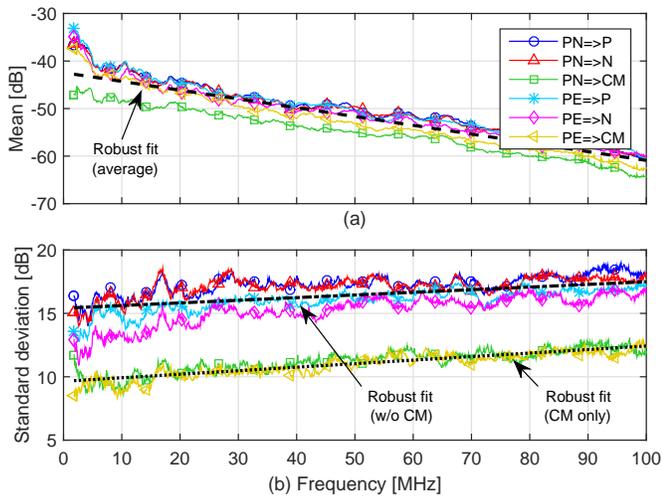}
\caption{Mean (a) and standard deviation (b) along frequency for the best normal fit of the measured amplitudes in dB scale for all the considered transmitter-receiver combinations. The corresponding robust fit is also shown.}
\label{fig:MuSigma}
\end{figure}

From the analysis of the measurements in \cite{ETSI1}, the following properties have been found. The amplitude of $H(f)$ is log-normally distributed, hence $A_{dB}(f)$ is normally distributed, while the phase is uniformly distributed in $[-\pi,\pi)$, at any frequency and for each individual transmitting-receiving mode combination. This confirms the findings about SISO channels in \cite{Galli_09} and in \cite{TCOMp1}. In particular, the mean ($\mu_{A_{dB}}$) and the standard deviation ($\sigma_{A_{dB}}$) for the best normal fit of the CFR amplitudes in dB are reported in Fig.~\ref{fig:MuSigma} as a function of frequency and for all the considered transmitter-receiver mode combinations in the 1.8--100 MHz band.

\begin{table}[t]
\renewcommand{\arraystretch}{\tabstretch}
\caption{Robust regression fit coefficients for the mean and standard deviation parameters of the amplitude in {\normalfont dB} scale.}
\label{tab:RobFitCoef}
\centering
\begin{tabular}{c|c|c}
   \multirow{2}{*}{Parameter} & slope & $y$-intercept \\
    & (dB/GHz) & (dB) \\
  \hline
  $\mu_{A_{dB}}$ (unique) & $-184.68$ & $-42.44$ \\
  $\sigma_{A_{dB}}$ (w/o CM) & $20.86$ & $15.41$ \\
  $\sigma_{A_{dB}}$ (CM only) & $27.80$ & $9.64$ \\
\end{tabular}
\end{table}
It can be noted as the mean exhibits approximately the same decreasing profile along frequency for all the different transmitting-receiving mode combinations. Only the $\mu_{A_{dB}}$ value related to the PN$\Rightarrow$CM mode exhibits a slightly different trend, since it assumes lower values, especially at low frequencies. Thus, all the combinations can be approximated with a unique robust regression fit profile corresponding to the average trend. The robust fit coefficients for the mean profile are listed in Table~\ref{tab:RobFitCoef}. 

The standard deviation behavior, instead, is characterized by a slightly increasing profile and it is more spread among the different mode combinations. The lowest $\sigma_{A_{dB}}$ profiles are obtained for the combinations with the CM configuration at the receiver side. Hence, two different robust fit trends can be identified, namely one for the modes that do not consider the CM (labeled with w/o CM) and one for the modes that consider only the combinations with the CM (CM only). The robust fit coefficients for the standard deviation profiles are listed in Table~\ref{tab:RobFitCoef}.

Furthermore, the correlation properties of the considered database of MIMO channel measurements have been assessed. In particular, the overall normalized covariance matrix $\mathbf{R}_{H_{dB}}$ of the reshaped channel vector $\widetilde{\mathbf{H}}_{dB}$, obtained as described in Section~\ref{ssec:ThForm}, as well as the normalized covariance matrix between the amplitude (in dB scale) and the phase of $\widetilde{\mathbf{H}}$, i.e. $\mathbf{R}_{A_{dB},\varphi}$, are computed by considering the overall measurements database discussed in Section~\ref{ssec:MisCamp}. The former is approximately a real quantity, since the imaginary part shows a maximum value of $0.29$ and an average value of $0.033$, with only the $0.02$ \% of coefficients exceeding $0.2$, which is assumed sufficiently low to be neglected. This means that the imaginary part of $\mathbf{R}_{H_{dB}}$, expressed as in (\ref{eq:CovDep}), is nearly zero. The normalized covariance matrix $\mathbf{R}_{A_{dB},\varphi}$, instead, has a maximum and average value of $0.41$ and $0.046$, respectively. Despite some high values, only the $0.3$ \% of them exceed the quantity of $0.2$, indicating that the relationship between the amplitude in dB and the phase can be neglected. Therefore, $A_{dB}$ and $\varphi$ can be assumed as uncorrelated and are treated as independent in order to further simplify the modeling process. Thus, from the experimental evidence, they can be independently generated according to the corresponding statistical distribution and with the proper correlation level, as it will be clarified in the next section.

\begin{figure}[t]
\centering
\includegraphics[width=\figurewidth]{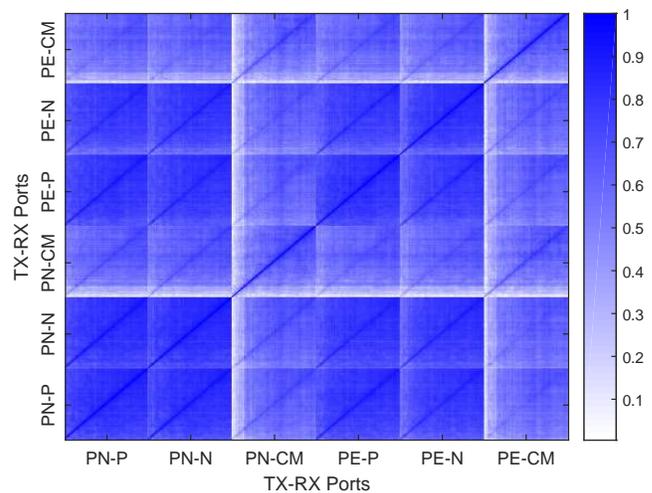}
\caption{Amplitude (in dB) normalized covariance matrix of the MIMO CFR measurements for all the considered transmitter-receiver mode combinations. Each depicted block identifies the frequency correlation exhibited among the corresponding modes pair.}
\label{fig:CorrMtxAmp}
\end{figure}

The absolute value of the amplitude (in dB) normalized covariance matrix $\mathbf{R}_{A_{dB}}$ is depicted in Fig.~\ref{fig:CorrMtxAmp}.
It can be noted as $\mathbf{R}_{A_{dB}}$ is a block matrix, where each block is referred to as $\mathbf{R}_{A_{dB}}^{i,j}$, with $i,j=1,\dots,6$ being the indexes of the considered mode combination. Each distinct point of a specific block refers to the correlation coefficient among the CFRs at two given frequencies associated to the ($i$,$j$) transmitting-receiving mode pairs. For example, the main diagonal blocks refer to the frequency correlation matrix of the same transmitting-receiving mode pair. Contrariwise, the out-of-diagonal blocks concern all the possible mode cross-combinations. Moreover, the block matrix depicted in Fig.~\ref{fig:CorrMtxAmp} is symmetric by definition.

The matrix consists of 36 blocks, with dimension $N_f \times N_f$ and $N_f=1588$, corresponding to all the possible combinations of transmitting and receiving mode pairs. Since the $2\times3$ MIMO channel is considered, the number of possible pairs is 6, leading to the 36 combinations. A high correlation level can be noticed almost everywhere, with lower values for low frequencies, especially for the combinations with the CM. This correlation behavior is due to the cross-talk phenomenon, which becomes prominent at high frequencies.

\begin{figure}[t]
\centering
\includegraphics[width=\figurewidth]{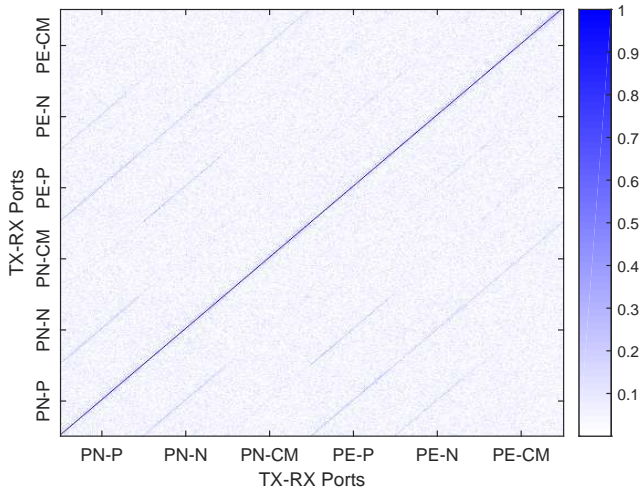}
\caption{Phase normalized covariance matrix of the MIMO CFR measurements for all the considered transmitter-receiver mode combinations. Each depicted block identifies the frequency correlation exhibited among the corresponding modes pair.}
\label{fig:CorrMtxPhi}
\end{figure}

Concerning the phase, the absolute value of the normalized covariance matrix $\mathbf{R}_{\varphi}$ is depicted in Fig.~\ref{fig:CorrMtxPhi}.
Obviously, the matrix is structured as the amplitude normalized covariance matrix in Fig.~\ref{fig:CorrMtxAmp}. However, in this case, high correlation values are noticeable only in proximity of the main diagonal, with some minor values on the main diagonal of the out-of-diagonal blocks. This is due to the large amount of branches that characterize a typical indoor PLC network topology, which cause multipath propagation. The resulting signal replicas combine at the receiver leading to uniformly and uncorrelated CFR phase values between distinct frequencies and regardless the considered MIMO modes combination, also due to the high considered frequencies. Thus, only the matrix $\mathbf{R}_{A_{dB}}$ significantly contributes to the overall covariance matrix. However, the high $\mathbf{R}_{\varphi}$  values noticeable on the main diagonal of the out-of-diagonal blocks suggest a strong correlation among the phase values of the different MIMO modes at the same frequency. This relationship is exploited as discussed in Section~\ref{ssec:PhiGen}.

In the following, it will be shown that both matrices are needed in the not-fully synthetic modeling approach. Instead, in the completely synthetic approach the amplitude covariance matrix is synthesized with a small set of parameters, while the phase covariance matrix is not needed.

\section{Synthetic Modeling Strategy}
\label{sec:ModelDescript}
This section details the procedure adopted to generate a set of simulated channels that show the same statistical properties of the actual measurements described in Section~\ref{ssec:ExpChPropMODEL}. The aim is to propose a practical implementation of the theoretical model discussed in Section~\ref{ssec:ThForm}. The proposed synthetic modeling strategy is then validated via numerical results by comparing simulated and experimental channels in terms of the main statistical metrics and channel capacity, i.e. the maximum achievable rate defined by Shannon.

A first direct procedure, named not-fully synthetic, that can be used for the numerical channel generation is based on the strategy discussed in \cite[Sec.~3.1.1]{IEICE_14} for the SISO case. As detailed, amplitude and phase are independently generated. Concerning the amplitudes (in dB scale), an equivalent set of independent normal RVs is generated and then correlated according to the experimental covariance matrix via a simple multiplication. The strategy for the phase generation is a bit more complicated due to its uniform statistics. However, exploiting the Pearson (linear correlation) and Spearman (rank correlation) relationship, as detailed in \cite[Sec.~3.1.1]{IEICE_14}, an equivalent set of uniform correlated RVs can be obtained. Although this strategy is straightforward, the experimental amplitude and phase covariance matrices need to be known. Hence, a very large number of coefficients (equal to $36\times1588^2=90,782,784$ in our case) must be provided in order to enable the model implementation. To overcome this issue, a fully synthetic strategy is proposed in the following, which allows to reduce the parameters as well as the implementation replicability. The amplitude and the phase generation procedures are separately considered since, in practice, they are independently generated.

\subsection{Amplitude Generation Procedure}
\label{ssec:AmpGen}
As discussed in Section~\ref{ssec:ExpChPropMODEL}, the CFR amplitudes are log-normally distributed, i.e. the dB version exhibits a normal distribution. The generation of a 3D $N_R \times N_T \times N_f$ MIMO matrix $\mathbf{A}_{dB}$ with correlated normal entries is a quite easy process. According to the procedure briefly discussed in \cite[Sec.~3.1.1]{IEICE_14} for the SISO case, a vector of normally distributed and independent RVs with zero mean and unit variance is considered first, i.e. the vector $\mathbf{N}\sim\mathcal{N}(0,1)$ of size $M$. Then, the vector of normal and correlated RVs is computed as \cite{papoulis2002probability}
\begin{equation}
\label{eq:AmpGen}
  \widetilde{\mathbf{A}}_{dB}=\mathbf{Q}_{A_{dB}}^{1/2}\mathbf{N}+\boldsymbol{\mu}_{A_{dB}},
\end{equation}
where $\{\cdot\}^{1/2}$ stands for the square root and $\boldsymbol{\mu}_{A_{dB}}=E[\widetilde{\mathbf{A}}_{dB}]$ identifies the vector whose entries are the amplitude mean parameter profiles depicted in Fig.~\ref{fig:MuSigma}a. The amplitude covariance matrix $\mathbf{Q}_{A_{dB}}$ is also obtained experimentally as discussed in Section~\ref{ssec:ThForm}. Finally, the desired 3D MIMO amplitude matrix $\mathbf{A}_{dB}$ can be easily reconstructed by reshaping the amplitudes vector $\widetilde{\mathbf{A}}_{dB}$.

In order to limit the information amount needed for the implementation, the following procedure is proposed. It uses approximated, yet analytic, profiles for the mean, variance and covariance matrix.

The mean and standard deviation profiles can be approximated via a robust regression fit with the parameters in Table~\ref{tab:RobFitCoef}, as shown in Fig.~\ref{fig:MuSigma}. In particular, due to the similarity between the different $\mu_{A_{dB}}$ profiles, a single robust fit, corresponding to the average among all the profiles, is considered. Instead, for the standard deviation two distinct behaviors can be identified, so that they are approximated via two robust fit profiles, i.e., one for the mode combinations that do not consider the CM and one for the combinations with the CM only.

Also the MIMO amplitude covariance matrix $\mathbf{Q}_{A_{dB}}$ of size $M \times M$ can be approximated. Indeed, although not immediately apparent and rather challenging to be identified, an analytic method to reconstruct the covariance matrix with few parameters can be used. The simplification method considers the corresponding normalized covariance matrix $\mathbf{R}_{A_{dB}}$ depicted in Fig.~\ref{fig:CorrMtxAmp}. As intuition suggests, each out-of-diagonal block of size $N_f \times N_f$, which represents the mutual correlation among the considered modes, must be somehow related to the corresponding diagonal blocks that show the auto-correlation of each considered mode combination. This is confirmed by the experimental analysis that shows as the out-of-diagonal block in position ($i,j$), i.e. $\mathbf{R}_{A_{dB}}^{i,j}$, where $i,j=1,...,6$ stand for the corresponding transmitting and receiving mode combinations, is approximately related to the corresponding diagonal blocks in position ($i,i$) and ($j,j$) according to
\begin{equation}
\label{eq:CorrMtxGen}
  \mathbf{R}_{A_{dB}}^{i,j}=\frac{\mathbf{R}_{A_{dB}}^{i,i}\mathbf{R}_{A_{dB}}^{j,j}/N_f+\mathbf{R}_{A_{dB}}^{i,i}\circ\mathbf{R}_{A_{dB}}^{j,j}}{2},
\end{equation}
where $\circ$ denotes the Hadamard product, while the apex $\{\cdot\}^{i,j}$ refers to the considered modes combination, or block position.

The result above can be motivated from the findings in Section~\ref{ssec:ExpChPropMODEL} and examining the matrix in Fig.~\ref{fig:CorrMtxAmp}. Firstly, the MIMO amplitudes covariance matrix is symmetric. Then, also the blocks on the main diagonal must be symmetric. However, the individual out-of-diagonal blocks in the lower (or upper) triangular part of the overall matrix are no more symmetric and all their coefficients need to be specified. It has been observed that the out-of-diagonal blocks can be well approximated via the average of the two terms appearing in the numerator of (\ref{eq:CorrMtxGen}). The first term is the matrix product among the corresponding diagonal blocks, divided by the number of rows (or columns) in a block, i.e. $N_f$. Hence, each element of the resulting matrix is the average of the corresponding row-by-column products of the two considered diagonal blocks. However, this first quantity misses to consider the high correlation values noticeable in the proximity of the main diagonal of each block. Thus, the second quantity in (\ref{eq:CorrMtxGen}) is considered and consists of the Hadamard product among the corresponding diagonal blocks. 

In summary, all the points in the lower (or upper) triangular part of the overall matrix $\mathbf{R}_{A_{dB}}$, namely $(6/2+15)N_f^2$ points, can be reconstructed from the lower (or upper) triangular part of the corresponding diagonal blocks, having a total of $3N_f^2$ points. Thus, the number of required coefficients is reduced by $6$ times. Then, the covariance matrix $\mathbf{Q}_{A_{dB}}$, required for the amplitudes generation process according to (\ref{eq:AmpGen}), is computed from $\mathbf{R}_{A_{dB}}$ as
\begin{equation}
\label{eq:QfromR}
\mathbf{Q}_{A_{dB}}=\mathbf{R}_{A_{dB}}\circ(\boldsymbol{\sigma}_{A_{dB}}\boldsymbol{\sigma}^T_{A_{dB}}),
\end{equation}
where the approximated standard deviation profiles of the different mode combinations reported in Fig.~\ref{fig:MuSigma}b are grouped in an unique vector $\boldsymbol{\sigma}_{A_{dB}}$, similarly to the vector of mean profiles $\boldsymbol{\mu}_{A_{dB}}$. 

\begin{figure}[t]
\centering
\includegraphics[width=\figurewidth]{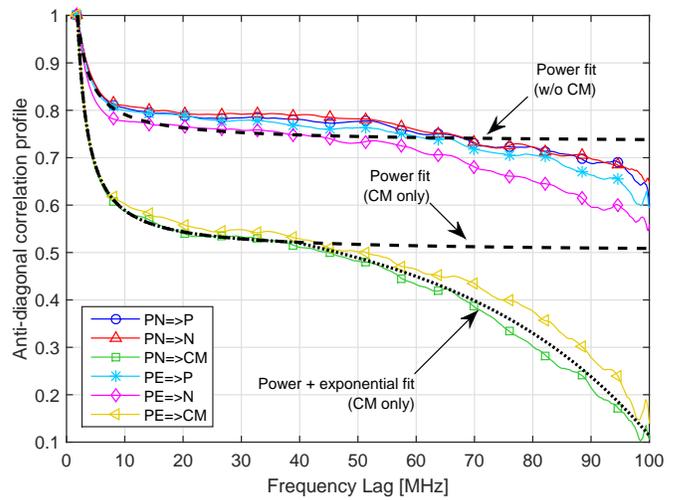}
\caption{Anti-diagonal profiles for each diagonal block related to the different MIMO transmitter-receiver combinations. The power fit is also depicted.}
\label{fig:AntiDiagProf}
\end{figure}

A further simplification is obtained by providing a method to reconstruct the diagonal blocks that represent the frequency correlation for a given modes combination. Strictly speaking, the correlation of the CFR at two frequencies does not depend only on the lag between frequencies (which corresponds to an unstationary CFR), as it can be noted in Fig.~\ref{fig:CorrMtxAmp}. However, it has been observed that a good approximation still holds if each diagonal block is approximated by a matrix that depends on the frequency lag only, resulting in a diagonally striped matrix (Toeplitz matrix). Each stripe has a constant value given by the average of the corresponding out-of-diagonal elements of the considered experimental MIMO matrix diagonal block. The anti-diagonal profile for each approximated diagonal block is depicted in Fig.~\ref{fig:AntiDiagProf} as a function of the frequency lag.

Similarly to the standard deviation, two different anti-diagonal profiles can be identified, depending on whether the CM is considered or not. The two families of trends are averaged and approximated through a fitting equation that exhibits a power frequency dependence according to the relation
\begin{equation}
\label{eq:PowFit}
  y(f)=af^b+c,
\end{equation}
where $a$, $b$ and $c$ are constant coefficients, while $y(f)$ identifies the anti-diagonal profile along frequency. The coefficients for both the profiles, referred to as power fits, are listed in Table~\ref{tab:PowFitCoef}.
\begin{table}[t]
\renewcommand{\arraystretch}{\tabstretch}
\caption{Power and exponential fit factors for all the combinations.}
\label{tab:PowFitCoef}
\centering
\begin{tabular}{c|c|c|c|c}
  Combinations & Type & $a$ & $b$ & $c$ \\
  \hline
  w/o CM & Power & $0.133 \times 10^6$ & $-0.906$ & $0.731$ \\
  \multirow{2}{*}{CM only} & Power & $1.679 \times 10^6$ & $-1.040$ & $0.501$ \\
   & Exp. & $-0.022$ & $0.031 \times 10^{-6}$ & $0.072$ \\
\end{tabular}
\end{table}

Since one of the most common statistical metrics that characterizes the channel, namely the CB at level 0.9, is given by the frequency at which the absolute value of the CFR correlation falls below 0.9 times its maximum in zero, it is fundamental to accurately approximate the correlation values for low frequency lags. Hence, the best power fits depicted in Fig.~\ref{fig:AntiDiagProf} (dashed lines), and their corresponding coefficients reported in Table~\ref{tab:PowFitCoef}, are computed focusing on the behavior up to 40 MHz. Although the power fit represents a good approximation for the mode combinations without the CM, a discrepancy exists for the combinations with the CM only, for large frequency lags. However, this difference can be neglected since, as said, mainly the lower frequency lags concur to provide a certain CB level.

If a better fit is desired for the CM combinations, a mixture of the power fit in (\ref{eq:PowFit}) plus an exponential fit of the form $y(f)=ae^{bf}+c$ can be adopted in order to improve the trend approximation, as shown in Fig.~\ref{fig:AntiDiagProf} (dotted line). The best exponential fit is computed referring to the frequencies above 40 MHz of the vertically shifted profile so that the first element (at 40 MHz) is zero. This way, it is possible to directly add the resulting fit to the power fit. The overall fit is depicted in Fig.~\ref{fig:AntiDiagProf}, while the exponential fit factors are listed in Table~\ref{tab:PowFitCoef}.

Thus, with the proposed strategy, it is possible to approximate the experimental MIMO amplitude covariance matrix of extremely large size (more than 90 million values), with an extremely small number of coefficients (equal to 12: 6 for the diagonal blocks and 6 for the mean and standard deviation profiles).

\subsection{Phase Generation Procedure}
\label{ssec:PhiGen}
The generation of a set of uniformly distributed RVs, with a given target normalized covariance matrix $\mathbf{R}_\varphi$, can be performed adopting the same procedure discussed in \cite[Sec.~3.1.1]{IEICE_14} for the SISO case, as it is done for the not-fully synthetic model. The difference is that the reshaped version of the 3D MIMO phase matrix must be considered for the computation. Then, the relation among the well known Pearson, or linear, correlation and the Spearman, or rank, correlation \cite{Gili_11} can be exploited as summarized in \cite[Sec.~3.1.1]{IEICE_14}. However, as discussed for the amplitude generation process in Section~\ref{ssec:AmpGen}, all the coefficients of the matrix $\mathbf{R}_\varphi$ need to be known. Hence, an approximated procedure is desirable.

Furthermore, although this approach is capable to reproduce the phase frequency and MIMO modes correlation exhibited by the overall set of simulated channel, it somewhat fails to represent the correct phase frequency correlation for an individual simulated realization. This inaccuracy translates into a difference in terms of RMS-DS and CB values among experimental and simulated channels, especially for the MIMO case, as noticeable in Table~\ref{tab:AVGvalMIMOmodel}. This is since, as demonstrated in \cite{TCOMp1}, the RMS-DS of each CFR is related to the corresponding unwrapped phase slope, which represents the phase correlation at different frequencies. However, this relationship can be exploited in order to generate the correct individual, i.e. realization by realization, phase frequency correlation by imposing a specific unwrapped phase behavior. The proposed synthetic model exploits this dependency for the phase generation procedure, as it is discussed in the following.

\begin{figure}[t]
\centering
\includegraphics[width=\figurewidth]{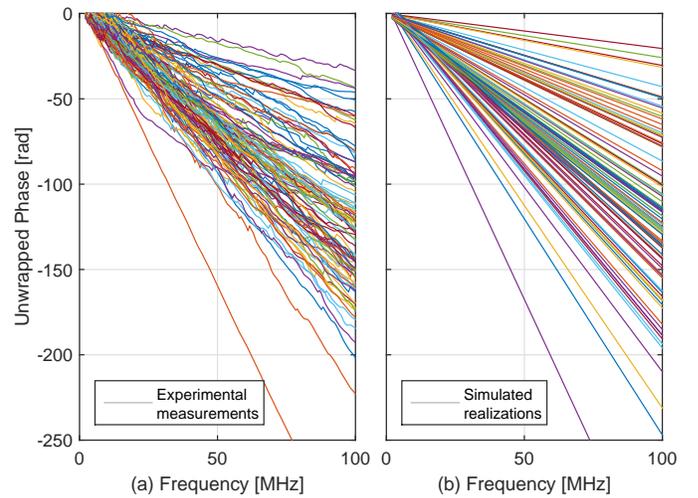}
\caption{Subset of 100 unwrapped phase profiles along frequency for the measurements (a) and the simulated channels (b), regardless the considered modes combination.}
\label{fig:UnwrapPhase}
\end{figure}

The unwrapped phase is computed from the conventional phase in $[-\pi,\pi)$ by adding $\pm2\pi$ if the phase jump between two consecutive frequency samples is greater than, or equal to, $\pm\pi$. A subset of 100 experimental unwrapped phase profiles, randomly chosen among the different MIMO transmitting-receiving mode combinations, is shown in Fig.~\ref{fig:UnwrapPhase}a. Indeed, as shown in Fig.~\ref{fig:CorrMtxPhi} and discussed in Section~\ref{ssec:ExpChPropMODEL}, only a marginal difference exists among the various MIMO modes, which is therefore neglected for the sake of the model simplicity.

As it can be noted, the unwrapped phase exhibits a linearly decreasing trend along frequency, with a different slope for each realization. Hence, the synthetic modeling strategy approximates each profile via the robust regression fit slope and neglecting the $y$-intercept contribution. This assumption works well because, since the starting frequency is not zero, when the simulated unwrapped phase trends are converted back to the conventional phase in $[-\pi,\pi)$, an uniform distribution is still obtained at each frequency in the 1.8--100 MHz band. This is due to the unwrapped phase slope variability and has also been proven by experimental evidence. The best maximum likelihood fit for the unwrapped phase slope, with opposite sign, turns out to be the generalized extreme value (GEV) distribution with parameters $\xi=-0.08$, $\mu_\varphi=1.133 \times 10^{-6}$ and $\sigma_\varphi=5.323 \times 10^{-7}$, namely the shape, location and scale parameters, respectively.

Thus, a set of simulated phase profiles, statistically equivalent to the measured ones, can be generated relying on different unwrapped phase trends having a slope that exhibits a GEV distribution with the above mentioned parameters. The same profile is adopted for all the MIMO modes of each realization, since, as discussed in Section~\ref{ssec:ExpChPropMODEL} and noticeable in Fig.~\ref{fig:CorrMtxPhi}, a significant correlation is present among them. A reduced set of 100 simulated unwrapped phase profiles, chosen at random among the various MIMO modes, is depicted in Fig.~\ref{fig:UnwrapPhase}b. The emulated trends show a similar behavior w.r.t. the experimental ones. This strategy is able to provide phase realizations with the right frequency correlation, hence resulting into correct RMS-DS and CB average values, as demonstrated by the numerical results discussed in Section~\ref{sec:Results}.

\subsection{Summary of the Overall Channel Generation Procedure}
A summary of the proposed channel modeling strategy is herein reported for clarity. According to the formulation in (\ref{eq:CFRdB}), a set of simulated channel realizations, which show the same statistical properties of a database of channel measurements, can be generated by modeling the second order statistics of the CFRs amplitude and phase. This is since, from the considerations in Section~\ref{ssec:ExpChPropMODEL}, the CFR amplitude and phase can be assumed and generated as independent.

The normally distributed and correlated amplitudes, in dB scale, are generated from uncorrelated normal random variables according to (\ref{eq:AmpGen}). The mean parameter profiles, reorganized in the reshaped mean vector $\boldsymbol{\mu}_{A_{dB}}$, are approximated by the robust fit trend with the coefficients listed in Table~\ref{tab:RobFitCoef}. Instead, the amplitudes MIMO covariance matrix $\mathbf{Q}_{A_{dB}}$ is reconstructed from the normalized covariance matrix $\mathbf{R}_{A_{dB}}$ as in (\ref{eq:QfromR}). The standard deviation parameter profiles, grouped in the reshaped vector $\boldsymbol{\sigma}_{A_{dB}}$, are approximated via the robust fit trends with the parameters listed in Table~\ref{tab:RobFitCoef}. Then, $\mathbf{R}_{A_{dB}}$ is generated according to the following steps. Firstly, the diagonal blocks of size $N_f \times N_f$ are approximated with diagonally striped matrices. Each diagonal stripe has identical elements along it. The stripes constant value is obtained from the anti-diagonal profile, reproduced according to (\ref{eq:PowFit}) and with the coefficients in Table~\ref{tab:PowFitCoef}, for the corresponding modes combination. Secondly, the out-of-diagonal blocks are computed from the diagonal blocks via (\ref{eq:CorrMtxGen}).

The uniformly distributed and frequency dependent phases are generated exploiting the linearly decreasing behavior of the unwrapped phase profile, as discussed in Section~\ref{ssec:PhiGen}. In particular, a set of unwrapped phase profiles are obtained via robust fits with a slope coefficient, inverted in sign, that has a GEV distribution with the parameters specified in Section~\ref{ssec:PhiGen}. The same profile is adopted for the different MIMO modes of each realization. Then, the phase samples are reconstructed by restoring the periodicity.

This generation procedure for the $2\times3$ MIMO PLC channel, named synthetic channel modeling, requires 15 parameters, 12 for the amplitude generation and 3 for the phase generation. On the contrary, the not-fully synthetic approach requires the knowledge of the overall measured amplitude and phase covariance matrices, which have extremely large size.

\section{Numerical Results}
\label{sec:Results}
To validate the synthetic modeling strategy detailed in Section~\ref{sec:ModelDescript}, the statistical metrics, the channel capacity, and the condition number ($\kappa$) defined as in \cite[Eq.~25]{CortesJSAC}, are herein assessed considering the overall database of measurements and all the simulated channel realizations. In order to provide a scalar value, both the average value ($\mu_x$) of the metrics, computed among the realizations and between all the considered MIMO modes, and their average standard deviation ($\sigma_x$), computed over the realizations and averaged among the MIMO modes, are considered. The subscript $x$ refers to the corresponding metric. When the CB is considered, the resolution in frequency ($\Delta f$) has to be sufficiently high in order to provide fair CB values. In particular, the considered measurements have a frequency resolution limited to $\Delta f=62.5$ kHz, due to the deployed instrumentation. However, $\Delta f$ is small enough to provide proper CB values, which is in the order of 200--300 kHz as discussed in the following.

The MIMO channel capacity, under the colored and spatially correlated Gaussian noise assumption, is also analyzed and calculated according to \cite[Eq.~6]{MIMOcapacity}. Indeed, it strongly depends on the CFR properties, such as the frequency and MIMO modes correlation, especially within the MIMO context, since it is computed through a determinant operator. Thus, it is a measure of the ability of the model to match the experimental MIMO channels. Furthermore, the capacity is often used to evaluate physical layer algorithms performance.

In practice, the results obtained considering the 353 MIMO measurements described in Section~\ref{ssec:MisCamp} are compared to those achieved by a set of 353 simulated channel realizations. Clearly, an arbitrarily large number of channels can be generated. In particular, the proposed synthetic model is tested for the $2\times3$ MIMO scheme in Section~\ref{ssec:2x3MIMO} and for a subset of MIMO modes in Section~\ref{ssec:2x2MIMO_SISO}, namely $2\times2$ MIMO and SISO. As a further comparison, the not-fully synthetic procedure is also considered for the $2\times3$ MIMO case.

\subsection{$2\times3$ MIMO Scheme}
\label{ssec:2x3MIMO}
In this paragraph, both the not-fully synthetic and the synthetic models are considered. The former, which is the MIMO extension of the initial idea proposed in \cite{IEICE_14} for the SISO scheme, is considered only for comparison purposes.

A first comparison is made in terms of ACG, RMS-DS, CB, and condition number $\kappa$, which is a measure of the correlation among the MIMO modes. The average and standard deviation values for the $2\times3$ MIMO scheme are listed in Table~\ref{tab:AVGvalMIMOmodel}. The reported experimental results differ from those displayed in \cite[Tab.~3]{PaganiJSAC}, although the same database is considered, since a different methodology is used for the statistical metrics computation. The presented results are computed according to the same procedure detailed in \cite{TCOMp1}.

The comparison of the values reported in Table~\ref{tab:AVGvalMIMOmodel} shows that both the not-fully synthetic and the synthetic channel exhibit ACG parameters ($\mu_{A_{dB}}$ and $\sigma_{A_{dB}}$) in agreement with the measured channels. However, the not-fully synthetic procedure leads to significant differences in terms of RMS-DS metric, and conversely of CB, and also in terms of $\kappa$ values. It has been proven that this high level of channel dispersion (low correlation) is mainly due to the considerable phase randomness. Indeed, the phase generation procedure proposed in \cite{IEICE_14}, which is adopted for the not-fully synthetic model, exploits the Pearson and Spearman relation to obtain a correlation matrix equivalent to the experimental one. This strategy fairly reproduces the overall phase behavior, but it provides approximately random phase values along frequency for each realization. This effect involves a great CFR variability, leading to a very high RMS-DS and a zero CB, hence below the frequencies spacing. Furthermore, the RMS-DS values are more spreaded w.r.t. those related to the measurements, as shown by the standard deviation $\sigma_\tau$.
\begin{table}[t]
\renewcommand{\arraystretch}{\tabstretch}
\caption{Main statistical metrics for experimental and simulated channels, generated with synthetic and not-fully synthetic methods. The $2\times3$ MIMO scheme is considered.}
\label{tab:AVGvalMIMOmodel}
\centering
\begin{tabular}{c|c|c|c|c}
  Metric & Parm. & Experim. & Synthetic & Not-fully Synt. \\
  \hline
  ACG & $\mu_G$ & $-42.30$ & $-43.07$ & $-41.08$ \\
  (dB) & $\sigma_G$ & $9.93$ & $12.53$ & $11.07$ \\
  \hline
  RMS-DS & $\mu_{\tau}$ & $0.350$ & $0.335$ & $2.702$ \\
  ($\mu$s) & $\sigma_{\tau}$ & $0.226$ & $0.052$ & $0.415$ \\
  \hline
  CB & $\mu_B$ & $293.22$ & $217.71$ & $0$ \\
  (kHz) & $\sigma_B$ & $324.30$ & $53.76$ & $0$ \\
  \hline
  $\kappa$ & $\mu_\kappa$ & $14.26$ & $14.70$ & $10.62$ \\
  (dB) & $\sigma_\kappa$ & $7.25$ & $6.64$ & $5.16$ \\
  \hline
  Capacity & $\mu_C$ & $1.53$ & $1.49$ & $1.61$ \\
  (Gbps) & $\sigma_C$ & $0.74$ & $0.68$ & $0.70$ \\
\end{tabular}
\end{table}

\begin{figure}[t]
\centering
\includegraphics[width=\figurewidth]{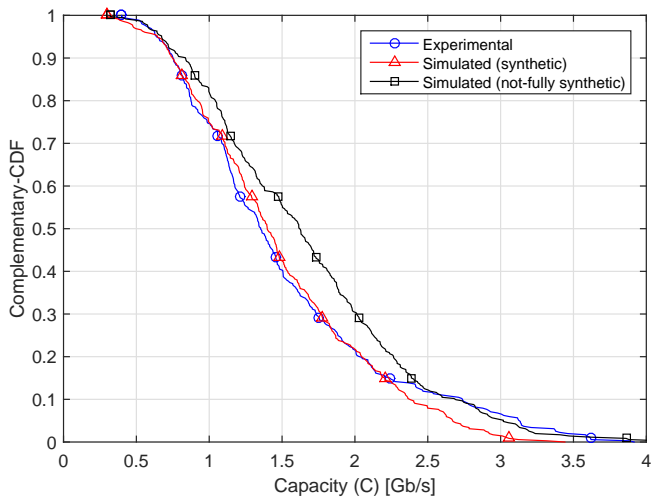}
\caption{Capacity CCDF comparison between the experimental and the simulated $2\times3$ MIMO channels in the 1.8--100 MHz band.}
\label{fig:CCDFcmp}
\end{figure}
Conversely, the matching among the measurements and the channels generated via the synthetic strategy is very high for the statistical metrics average value, especially in terms of RMS-DS. This is because the phase frequency dependence (correlation) is better modeled and the model underestimates the CB, which can be considered as a conservative result w.r.t. \cite{PaganiJSAC} that overestimates the CB. A discrepancy, instead, is noticeable in the standard deviation values related to the RMS-DS and CB metrics, as a result of trading off with modeling simplicity. Contrariwise, an almost perfect agreement is exhibited by the $\kappa$ metric, indicating the ability of the model to faithfully reproduce the experimental correlation among the MIMO modes.

As discussed at the beginning of this section, another quantity of great interest is represented by the channel capacity, which additionally provides an overall evaluation of the performance attainable by a communication system. The capacity complementary cumulative distribution function (CCDF) achieved with the experimental and the simulated $2\times3$ MIMO channels in the 1.8--100 MHz band is depicted in Fig.~\ref{fig:CCDFcmp}. The capacity is computed exploiting an optimal power allocation among the possible MIMO modes, with a total available power per frequency equal to the power spectral density (PSD) constraint specified in \cite{HPAV2}. Moreover, colored Gaussian noise, correlated among the MIMO ports, is considered. In particular, the noise PSD profiles along frequency taken from the measurement campaign in \cite{ETSI3} are used. Instead, the noise correlation between the three receiving ports is considered and generated as detailed in \cite{ISPLC_14}.
 
The experimental and both the simulated channels exhibit a similar capacity CCDF. The existing discrepancy might be due to the approximation of $A_{dB}$ and $\varphi$ as independent RVs. This simplification avoids the challenging generation of a pair of RVs that are statistically dependent with a certain joint statistical distribution, which is non-trivially detectable from the measurements.

The capacity obtained by the synthetic modeling strategy provides a better matching w.r.t. the not-fully synthetic procedure one, as highlighted by the capacity mean ($\mu_C$) and standard deviation ($\sigma_C$) values listed in Table~\ref{tab:AVGvalMIMOmodel} that concern the $2\times3$ MIMO channel. This is due to the improved capability to consider the proper phase correlation along frequency. The main differences are limited to the high rates tail where the model underestimates the capacity of favorable channels. Hence, the synthetic strategy provides conservative values.

\begin{figure}[t]
\centering
\includegraphics[width=\figurewidth]{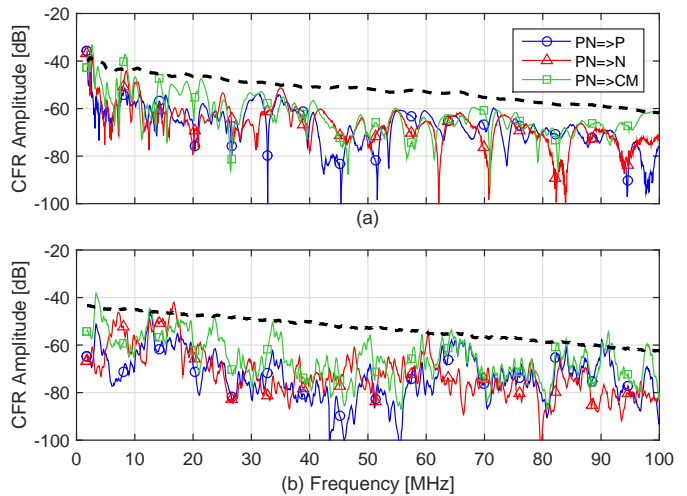}
\caption{Comparison among experimental (a) and simulated (b) $2\times3$ MIMO CFRs in the 1.8--100 MHz band. Only the receiving modes related to the PN transmission port are considered for simplicity. The average profile computed considering all the realizations and the receiving modes is also depicted.}
\label{fig:CFRcomp}
\end{figure}

As a last term of comparison, experimental and simulated $2\times3$ MIMO CFRs are depicted in Fig.~\ref{fig:CFRcomp}a and Fig.~\ref{fig:CFRcomp}b, respectively. For graphical purposes, only the PN transmitting port is considered in combination with all the three receiving ports.
It can be noted as the simulated MIMO CFR exhibits a good selectivity behavior along frequency, if compared to the measured one. The only difference is that the profile is a bit noisier, as confirmed by the lower CB values reported in Table~\ref{tab:AVGvalMIMOmodel}. As a further comparison, the average CFR profile, which is computed considering the overall measurements database along with all the receiving ports for the given PN transmission port, is also displayed (dashed black line).

\subsection{SISO and $2\times2$ MIMO Schemes}
\label{ssec:2x2MIMO_SISO}
This paragraph aims to validate the synthetic modeling strategy proposed in Section~\ref{sec:ModelDescript} also for a reduced set of MIMO modes, namely the $2\times2$ MIMO and the SISO transmitting schemes. Concerning the $2\times2$ MIMO scheme, both the \mbox{$\Delta$-style} transmitting ports are considered, limiting the reception to the first two star-style receiving ports, i.e. P and N. The SISO case considers the transmission among the PN \mbox{$\Delta$-style} transmitting port and the P star-style receiving port.

\begin{table}[t]
\renewcommand{\arraystretch}{\tabstretch}
\caption{Main statistical metrics of the $2\times2$ MIMO and the SISO schemes for experimental and simulated channels, generated according to the synthetic procedure.}
\label{tab:AVGvalSubset}
\centering
\begin{tabular}{c|c|c|c|c|c}
  \multirow{2}{*}{Metric} & \multirow{2}{*}{Parm.} & \multicolumn{2}{c|}{SISO} & \multicolumn{2}{|c}{$2\times2$ MIMO} \\
  & & Exp. & Sim. & Exp. & Sim. \\
  \hline
  ACG & $\mu_G$ & $-40.12$ & $-40.53$ & $-40.87$ & $-43.12$ \\
  (dB) & $\sigma_G$ & $12.79$ & $14.99$ & $12.00$ & $14.41$ \\
  \hline
  RMS-DS & $\mu_{\tau}$ & $0.353$ & $0.332$ & $0.357$ & $0.330$ \\
  ($\mu$s) & $\sigma_{\tau}$ & $0.266$ & $0.052$ & $0.249$ & $0.055$ \\
  \hline
  CB & $\mu_B$ & $342.50$ & $210.87$ & $316.65$ & $219.54$ \\
  (kHz) & $\sigma_B$ & $467.28$ & $51.01$ & $392.65$ & $56.20$ \\
  \hline
  $\kappa$ & $\mu_\kappa$ & $-$ & $-$ & $16.65$ & $18.74$ \\
  (dB) & $\sigma_\kappa$ & $-$ & $-$ & $8.81$ & $9.98$ \\
  \hline
  Capacity & $\mu_C$ & $0.76$ & $0.76$ & $1.35$ & $1.31$ \\
  (Gbps) & $\sigma_C$ & $0.43$ & $0.40$ & $0.76$ & $0.66$ \\
\end{tabular}
\end{table}
\begin{figure}[t]
\centering
\includegraphics[width=\figurewidth]{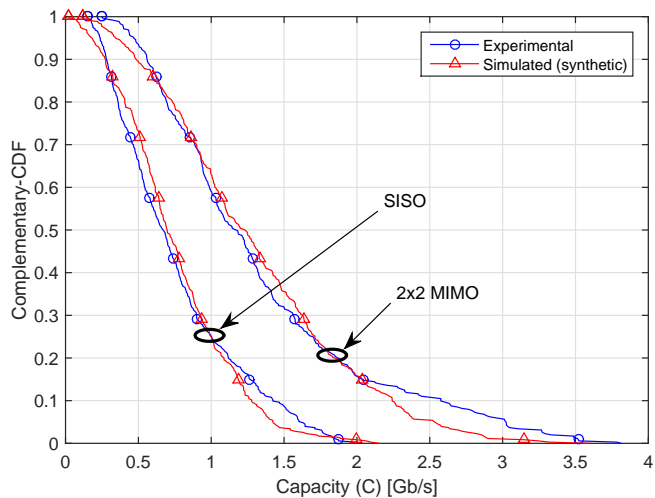}
\caption{Capacity CCDF comparison between experimental and simulated channels for the SISO and the $2\times2$ MIMO case in the 1.8--100 MHz band.}
\label{fig:CapCCDFsubset}
\end{figure}
The statistical metrics for the experimental and the simulated channels for both the $2\times2$ MIMO and SISO channels are listed in Table~\ref{tab:AVGvalSubset}. A good agreement is observed in terms of $\kappa$ and statistical metrics average value, with some differences only in terms of CB. This is mainly due to the approximation adopted for the amplitude covariance matrix anti-diagonal profile discussed in Section~\ref{ssec:AmpGen}. In addition, similar values are obtained also for the ACG average standard deviation, while the RMS-DS and CB experimental and simulated standard deviations are not so close. As discussed in Section~\ref{ssec:2x3MIMO}, this discrepancy is mainly related to the adopted simplifications. Nevertheless, a good performance replicability is obtained, as confirmed by the capacity CCDF.

The experimental and simulated channel capacity CCDF for both the schemes $2\times2$ MIMO and SISO is shown in Fig.~\ref{fig:CapCCDFsubset}. It can be noted as there is a good agreement between the performance achieved by the measurements and those obtained by the simulated channels generated according to the synthetic modeling strategy. In particular, an almost perfect matching is noticeable for the SISO transmission scheme, while some deviations are found for the $2\times2$ MIMO case when considering high capacity values.

\section{Conclusions}
\label{sec:conc}
Modeling the PLC channel is relevant to support the task of testing and developing communication solutions over power lines. A brief overview has been given concerning the main modeling approaches present in the literature, mostly limited to the SISO scenario. In this paper, an extremely compact and effective top-down modeling method has been proposed for the MIMO PLC channel, which has been derived from a pure phenomenological approach. It is based on the synthesis of the second order channel statistics observed in a real data set. The model is compact and requires few parameters to simulate an in-home $2\times3$ MIMO PLC channel in the band 1.8--100 MHz, i.e. the appropriate CFR amplitude and phase statistics and their modes and frequency correlation. The results have validated the proposed strategy as a statistically representative simulation tool, despite its simplicity.

A comparison to the not-fully synthetic procedure proposed in \cite[Sec.~3.1.1]{IEICE_14} for the SISO case, herein extended to the MIMO case, has also been made. It has been shown that, despite the extremely reduced set of parameters required by the synthetic model w.r.t. the not-fully synthetic model, a good matching with the experimental channel statistics and channel capacity distribution is achieved. 

The proposed synthetic channel model has been implemented in a channel software generator, using the numerical computing environment Matlab, that is available online \cite{OnlineScript}.

\newpage

\begin{IEEEbiography}[{\includegraphics[width=1in,height=1.25in,clip,keepaspectratio]{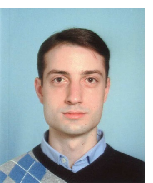}}]{Alberto Pittolo}
received the Bachelor of Science degree in electrical engineering (2009), the Master of Science degree (2012) in electrical and telecommunications engineering, with honors, and the PhD (2015) in industrial and information engineering from the University of Udine, Udine, Italy. His research interests and activities focus on the channel statistical characterization and on its numerical modeling, on the physical layer security investigation, and on the resource optimization in conjunction with the allocation algorithms, for both wireless and power line communications (PLC).
\end{IEEEbiography}

\begin{IEEEbiography}[{\includegraphics[width=1in,height=1.25in]{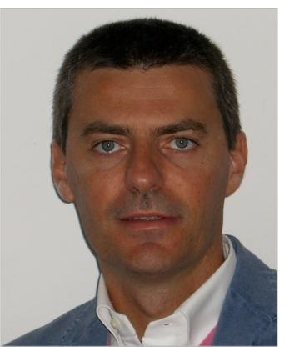}}]{Andrea M. Tonello}
(M00-SM12) received the Laurea degree (summa cum laude) and the PhD in electronics and telecommunications from the University of Padova, Italy, in 1996 and in 2003, respectively. From 1997 to 2002, he was with Bell Labs-Lucent Technologies, Whippany, NJ, USA, first as a Member of the Technical Staff. Then, he was promoted to Technical Manager and appointed to Managing Director of the Bell Labs Italy division. In 2003, he joined the University of Udine, Italy, where he became Aggregate Professor in 2005 and Associate Professor in 2014. He also founded WiTiKee, a spin-off company working on telecommunications for smart grids. Currently, he is Professor and Chair of the Embedded Communication Systems Group at the University of Klagenfurt, Austria. Dr. Tonello received several awards, including eight best paper awards, the Bell Labs Recognition of Excellence Award (1999), the Distinguished Visiting Fellowship from the Royal Academy of Engineering, U.K. (2010), the IEEE VTS Distinguished Lecturer Award (2011-2015), the Italian Full Professor Habilitation (2013). He is the Chair of the IEEE Communications Society Technical Committee on Power Line Communications. He serves as Associate Editor of IEEE Trans. on Communications and IEEE Access. He was the General Chair of IEEE ISPLC 2011 and IEEE SmartGridComm 2014, and TPC Co-Chair in a number of conferences.
\end{IEEEbiography}


\begin{thebibliography}{10}
\providecommand{\url}[1]{#1}
\csname url@samestyle\endcsname
\providecommand{\newblock}{\relax}
\providecommand{\bibinfo}[2]{#2}
\providecommand{\BIBentrySTDinterwordspacing}{\spaceskip=0pt\relax}
\providecommand{\BIBentryALTinterwordstretchfactor}{4}
\providecommand{\BIBentryALTinterwordspacing}{\spaceskip=\fontdimen2\font plus
\BIBentryALTinterwordstretchfactor\fontdimen3\font minus
  \fontdimen4\font\relax}
\providecommand{\BIBforeignlanguage}[2]{{%
\expandafter\ifx\csname l@#1\endcsname\relax
\typeout{** WARNING: IEEEtran.bst: No hyphenation pattern has been}%
\typeout{** loaded for the language `#1'. Using the pattern for}%
\typeout{** the default language instead.}%
\else
\language=\csname l@#1\endcsname
\fi
#2}}
\providecommand{\BIBdecl}{\relax}
\BIBdecl

\bibitem{MIMObook}
L.~Berger, A.~Schwager, P.~Pagani, and D.~Schneider, \emph{{MIMO Power Line
  Communications: Narrow and Broadband Standards, EMC, and Advanced
  Processing}}, ser. {Devices, Circuits, and Systems}.\hskip 1em plus 0.5em
  minus 0.4em\relax Taylor \& Francis, 2014.

\bibitem{HPAV2}
L.~Yonge, J.~Abad, K.~Afkhamie, L.~Guerrieri, S.~Katar, H.~Lioe, P.~Pagani,
  R.~Riva, D.~M. Schneider, and A.~Schwager, ``{An Overview of the HomePlug AV2
  Technology},'' \emph{{Journal of Electrical and Computer Engineering}}, vol.
  2013, pp. 1--20, 2013.

\bibitem{ITU_Ghn}
\BIBentryALTinterwordspacing
``{Unified High-Speed Wireline-Based Home Networking Transceivers -- System
  Architecture and Physical Layer Specification},'' {Recommendation ITU-T
  G.9960}, {ITU}, December 2010. [Online]. Available:
  \url{https://www.itu.int/rec/T-REC-G.9960/en}
\BIBentrySTDinterwordspacing

\bibitem{LampeTonello_book}
L.~Lampe, A.~M. Tonello, and T.~G. Swart, \emph{{Power Line Communications:
  Principles, Standards and Applications from Multimedia to Smart Grid}},
  2nd~ed.\hskip 1em plus 0.5em minus 0.4em\relax Chichester, UK: {John Wiley \&
  Sons}, 2015.

\bibitem{SmartGrid2030}
S.~Goel, S.~F. Bush, and D.~Bakken, Eds., \emph{{IEEE Vision for Smart Grid
  Communications: 2030 and Beyond}}.\hskip 1em plus 0.5em minus 0.4em\relax
  {IEEE Standard Association}, 2013.

\bibitem{Galli_06}
S.~Galli and T.~C. Banwell, ``{A Deterministic Frequency-Domain Model for the
  Indoor Power Line Transfer Function},'' \emph{{IEEE Journal on Selected Areas
  in Communications}}, vol.~24, no.~7, pp. 1304--1316, July 2006.

\bibitem{BotUpMod_p1}
A.~M. Tonello and F.~Versolatto, ``{Bottom-Up Statistical PLC Channel Modeling
  -- Part I: Random Topology Model and Efficient Transfer Function
  Computation},'' \emph{{IEEE Transactions on Power Delivery}}, vol.~26, no.~2,
  pp. 891--898, April 2011.

\bibitem{PLC_MPM}
M.~Zimmermann and K.~Dostert, ``{A Multipath Model for the Powerline
  Channel},'' \emph{{IEEE Transactions on Communications}}, vol.~50, no.~4, pp.
  553--559, April 2002.

\bibitem{Tonello_12}
A.~M. Tonello, F.~Versolatto, B.~B{\'e}jar, and S.~Zazo, ``{A Fitting Algorithm
  for Random Modeling the PLC Channel},'' \emph{{IEEE Transactions on Power
  Delivery}}, vol.~27, no.~3, pp. 1477--1484, June 2012.

\bibitem{ETSI1}
{ETSI TR 101 562-1 V 1.3.1}, ``{PowerLine Telecommunications (PLT); MIMO PLT;
  Part 1: Measurement Methods of MIMO PLT},'' {European Telecommunication
  Standardization Institute}, Tech. Rep., 2012.

\bibitem{ETSI3}
{ETSI TR 101 562-3 V 1.1.1}, ``{PowerLine Telecommunications (PLT); MIMO PLT;
  Part 3: Setup and Statistical Results of MIMO PLT Channel and Noise
  Measurements},'' {European Telecommunication Standardization Institute},
  Tech. Rep., 2012.

\bibitem{Versolatto_11}
F.~Versolatto and A.~M. Tonello, ``{An MTL Theory Approach for the Simulation
  of MIMO Power-Line Communication Channels},'' \emph{{IEEE Transactions on
  Power Delivery}}, vol.~26, no.~3, pp. 1710--1717, July 2011.

\bibitem{CortesJSAC}
J.~A. Corchado, J.~A. Cort\'es, F.~J. Ca{\~n}ete, and L.~D\'iez, ``{An
  MTL-Based Channel Model for Indoor Broadband MIMO Power Line
  Communications},'' \emph{{IEEE Journal on Selected Areas in Communications}},
  vol.~34, no.~7, pp. 2045--2055, July 2016.

\bibitem{Canova_10}
A.~Canova, N.~Benvenuto, and P.~Bisaglia, ``{Receivers for MIMO-PLC channels:
  Throughput comparison},'' in \emph{{Proc. of IEEE International Symposium on
  Power Line Communications and Its Applications (ISPLC)}}, March 2010, pp.
  114--119.

\bibitem{PaganiJSAC}
P.~Pagani and A.~Schwager, ``{A Statistical Model of the In-Home MIMO PLC
  Channel Based on European Field Measurements},'' \emph{{IEEE Journal on
  Selected Areas in Communications}}, vol.~34, no.~7, pp. 2033--2044, July
  2016.

\bibitem{IEICE_14}
A.~M. Tonello, A.~Pittolo, and M.~Girotto, ``{Power Line Communications:
  Understanding the Channel for Physical Layer Evolution Based on Filter Bank
  Modulation},'' \emph{{IEICE Transactions on Communications}}, vol. E97-B,
  no.~8, pp. 1494--1503, August 2014.

\bibitem{TCOMp1}
A.~M. Tonello, F.~Versolatto, and A.~Pittolo, ``{In-Home Power Line
  Communication Channel: Statistical Characterization},'' \emph{{IEEE
  Transactions on Communications}}, vol.~62, no.~6, pp. 2096--2106, June 2014.

\bibitem{Gili_11}
M.~Gili, D.~Maringer, and E.~Schumann, \emph{{Numerical Methods and
  Optimization in Finance}}.\hskip 1em plus 0.5em minus 0.4em\relax Access
  Online via Elsevier, 2011, ch. 7: Modeling dependencies, pp. 165--173.

\bibitem{ProperComplex}
F.~D. Neeser and J.~L. Massey, ``{Proper complex random processes with
  applications to information theory},'' \emph{{IEEE Transactions on
  Information Theory}}, vol.~39, no.~4, pp. 1293--1302, Jul 1993.

\bibitem{Galli_09}
S.~Galli, ``{A simplified model for the indoor power line channel},'' in
  \emph{{Proc. of IEEE International Symposium on Power Line Communications and
  Its Applications (ISPLC)}}, March 2009, pp. 13--19.

\bibitem{papoulis2002probability}
\BIBentryALTinterwordspacing
A.~Papoulis and S.~U. Pillai, \emph{{Probability, random variables, and
  stochastic processes}}, ser. {McGraw-Hill electrical and electronic
  engineering series}.\hskip 1em plus 0.5em minus 0.4em\relax McGraw-Hill,
  2002. [Online]. Available:
  \url{https://books.google.ie/books?id=YYwQAQAAIAAJ}
\BIBentrySTDinterwordspacing

\bibitem{MIMOcapacity}
S.~Krusevac, P.~Rapajic, and R.~A. Kennedy, ``{Channel capacity estimation for
  MIMO systems with correlated noise},'' in \emph{{IEEE Global
  Telecommunications Conference (GLOBECOM)}}, vol.~5, Dec 2005, pp. 2810--2816.

\bibitem{ISPLC_14}
A.~Pittolo, A.~M. Tonello, and F.~Versolatto, ``{Performance of MIMO PLC in
  measured channels affected by correlated noise},'' in \emph{{Proc. of 18th
  IEEE International Symposium on Power Line Communications and its
  Applications (ISPLC)}}, March 2014, pp. 261--265.

\bibitem{OnlineScript}
\BIBentryALTinterwordspacing
A.~Pittolo and A.~M. Tonello, ``{Synthetic MIMO PLC Channel Emulator
  Software},'' February 2017, 1st release. [Online]. Available:
  \url{http://www.andreatonello.com/openresearch}
\BIBentrySTDinterwordspacing

\end{thebibliography}
\end{document}